# Astro2020 Science White Paper

# The Formation and Evolution of Multiple Star Systems

**Thematic Areas:** ☐ Planetary Systems  ☒ Star and Planet Formation
☐ Formation and Evolution of Compact Objects  ☐ Cosmology and Fundamental Physics
☐ Stars and Stellar Evolution  ☐ Resolved Stellar Populations and their Environments
☐ Galaxy Evolution  ☐ Multi-Messenger Astronomy and Astrophysics

**Principal Author:**
Name: John J. Tobin
Institution: National Radio Astronomy Observatory
Email: jtobin@nrao.edu
Phone: 434-244-6815

**Co-authors:** (names and institutions)
Marina Kounkel (University of Western Washington), Stella Offner (University of Texas at Austin), Patrick Sheehan (National Radio Astronomy Observatory), Doug Johnstone (NRC - Herzberg), S. Thomas Megeath (University of Toledo), Kaitlin M. Kratter (University of Arizona), Ian Stephens (Harvard-Smithsonian Center for Astrophysics), Zhi-Yun Li (University of Virginia), Sarah Sadavoy (Harvard-Smithsonian Center for Astrophysics), Leslie Looney (University of Illinois), Joel Green (STSci), Rob Gutermuth (University of Massachusetts), Will Fischer (STSci), Michael M. Dunham (SUNY - Fredonia), Yao-Lun Yang (University of Texas)

**Abstract:**
　Significant advances have been made over the past decade in the characterization of multiple protostar systems, enabled by the Karl G. Jansky Very Large Array (VLA), high-resolution infrared observations with the *Hubble Space Telescope*, and ground-based facilities. To further understand the mechanism(s) of multiple star formation, a combination of statistics, high-angular resolution radio/millimeter continuum imaging, characterization of kinematic structure, magnetic fields via polarimetry, and comparison with numerical simulations are needed. Thus, understanding the origin of stellar multiplicity in different regimes of companion separation will soon be within reach. However, to overcome challenges that studies in this field are now confronted with, a range of new capabilities are required: a new millimeter/centimeter wave facility with 10 mas resolution at $\lambda$ =1 cm, space-based near to far-infrared observatories, continued development of low to high resolution spectroscopy on 3m to 10m class telescopes, and an ELT-class telescope with near to mid-infrared imaging/spectroscopic capability.

Multiplicity is a ubiquitous feature of stellar populations. Nearly half of nearby Solar-type stars are found in binary or higher-order multiple systems [1], and the typical separation of solar-mass multiple stars is ~50 au. The stellar multiplicity frequency increases with stellar mass; stars more massive than the Sun have a higher multiplicity fraction and stars less massive than the Sun have a lower multiplicity fraction, but still greater than 25% [2]. Stellar multiplicity is important for understanding a broad range of phenomena from supernova rates to the incidence of stable planetary systems. However, the origin and evolution of stellar multiplicity remains unclear, but significant progress is being made and can be made in the next decade.

The high frequency of multiplicity observed in field star and pre-main sequence populations [2] points to an origin of multiplicity during the star formation process. The largest mass reservoir is available during the protostellar phase, making it the most promising epoch for companion star formation to occur [3]. But the mechanism by which most multiple stars form remains unclear. Even though excellent statistics are typically available for field stars, the formation route for multiple systems cannot be determined from evolved stellar populations alone because they have undergone millions to billions of years of dynamical evolution.

There are two favored routes to explain the formation of multiple star systems: disk fragmentation due to gravitational instability [4, 5, 6] and turbulent fragmentation within the molecular cloud [7, 8]. Disk fragmentation preferentially produces close (<500 au) multiple systems and requires a large ($R_{disk}$~50 au) and massive ($\gtrsim 0.1 \times M_{protostar}$) rotationally-supported disk around the primary star. Turbulent fragmentation can result in the formation of both wide and close multiple systems. In this scenario, the initial protostars form with separations of ~1000 au, can migrate inward to separations of ~100 au within ~10 kyr [8], and remain widely separated, or drift further apart. Rotation of the protostellar cloud itself has also been suggested as a formation mechanism [9], but current observational evidence points more toward turbulence for the formation of wide companions [10, 11, 12]. Thus, there can be a large fraction of apparent wide companions early in the star formation process that may not be bound, but could still affect neighboring systems. Finally, magnetic fields also likely have an influence the formation of multiple systems by reducing the size of disks and/or reducing the number of fragments in a collapsing cloud [29]. In order to reveal the origins of stellar multiplicity and its effects on proto-planetary systems, multiplicity must be characterized early.

To fully understand the formation of multiple star systems, several challenges must be met. 1) Improved statistics on multiplicity throughout the protostellar phase are required to understand the trends in multiple star formation. 2) The mass range of protostellar multiples must be understood to relate protostellar multiplicity to field stars. 3) Separations approaching au scales must be examined in the protostellar phase to understand the limits of fragmentation and migration. 4) These observations must be compared to multi-scale numerical simulations that sample the relevant scales and physics.

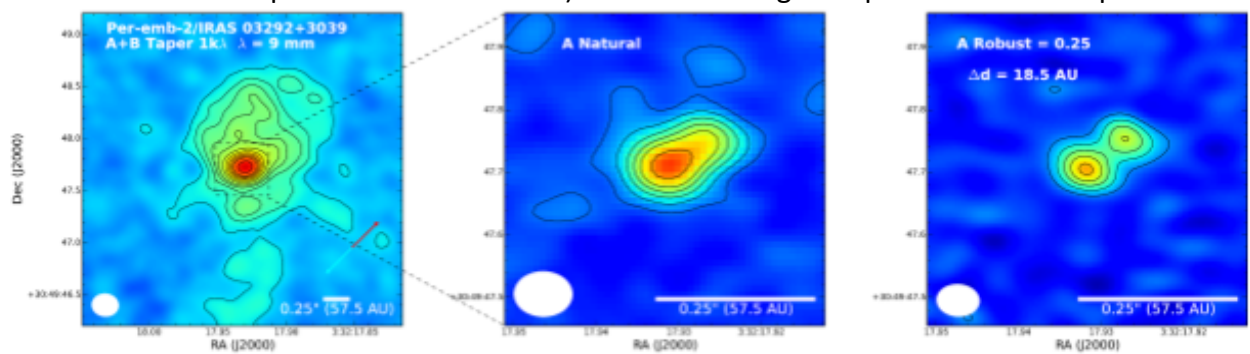

Figure 1. Multi-resolution images of a Class 0 proto-binary in the Perseus molecular cloud from the VLA. The left image shows the surrounding, massive disk of the protostar, while the middle and right images show the central part of the system at progressively higher resolution, resolving a very close binary.

**Surveys of the Youngest Protostellar Multiples**

Studies of multiplicity early in the protostellar phase, the Class 0 phase [13], are best conducted at millimeter to centimeter (mm/cm) wavelengths due to the protostar(s) being embedded within clouds having 10s to 100s $A_V$ of extinction. Early studies pointed to a high fraction of protostellar multiplicity [14, 15], but were hampered by a combination of small samples, low sensitivity, and poor angular resolution. The first survey to surmount many of these shortcomings was the VLA Nascent Disk and Multiplicity (VANDAM) Survey [11], carried out with the Karl G. Jansky Very Large Array (VLA). This survey observed all 82 protostars in the nearby (~300 pc) Perseus star forming region at ~20 au resolution. From these data, 18 multiple systems were identified with companion separations <500 au. The survey also detected a number of systems with separations >500 au as well as many hierarchical multiple systems. Some multiple systems had separations as small as ~20 au, near the limit of the spatial resolution. Figure 1 shows a very close multiple system, forming within a massive disk.

The separations of all companion stars detected in the VANDAM survey, are shown in Figure 2. Two features are obvious: 1) there appear to be two peaks, one at ~75 au and another >1000 au, and 2) the separation distribution is in excess of the field, except for separations between ~300 au and ~1000 au. It is argued that this bimodality results from both disk fragmentation (~75 au part) and turbulent fragmentation (>1000 au part) happening to produce the observed distribution [11]. The results of this survey clearly demonstrate that multiple star formation and fragmentation is a multi-scale process. A new survey of Orion has now been conducted with ~30 au resolution, with initial results showing a separation distribution similar to Perseus (Tobin et al. in prep.).

While this survey of Perseus represents a major advance, the number of sources was limited. Thus, it is clear that greater numbers are required and more than one star-forming region must be examined in order to understand the origin of multiplicity. The field solar-type sample sample [1] contains ~126 companion stars with separations between 1 au and 1000 au. In order to detect 126 protostellar companions (at a multiplicity frequency of ~20% [11]), at least 630 protostars must be observed. The 1 to 1000 au range of separations is where bound companion protostars are most like to be forming and feasible to observe. There are only ~156 Class 0 protostars in the entire Gould Belt [16]; thus, surveys of more distant, massive star forming regions are therefore required to substantially increase sample sizes and the range of primary masses sampled. The nearby regions typically only samples masses less than ~3 $M_{sun}$.

A challenge to observing larger numbers of protostars is that the most populous star forming regions are more distant. Orion is ~400 pc away and the VLA offers a best spatial resolution of ~30 au at 8 mm, which is just a bit better than the typical separation of field Solar-type stars. While spatial resolution decreases linearly with distance, the brightness of dust emission decreases as the distance-squared. For example, sources in Orion are ~1.33x more distant than those in Perseus, but are ~1.8x fainter and require ~3.2× more observing time to reach the same dust mass sensitivity (~120x more time at 1 kpc). As such, it is impractical to observe typical protostars (i.e., 1-3 $L_{sun}$ [16]) in more distant star forming regions, and studies of protostellar multiplicity in nearby (< 500 pc) clouds are reaching the limit of current capabilities.

The VLA also cannot observe protostars in nearby regions with significantly higher angular resolution. The Atacama Large Millimeter/submillimeter Array (ALMA) can achieve ~0.02" resolution at 1.3 mm, but this emission will primarily trace optically thick dust in protostellar disks and may prevent the detection of close companions. Thus, a facility capable of ~0.01" angular resolution and ~10x greater sensitivity operating between 3 mm to 2 cm, where dust is less opaque due to the longer wavelengths is required. Such a facility will open up the ability to characterize protostellar multiplicity toward regions out to ~1.5 kpc, like Cygnus-X, with the same level of detail currently available toward the nearby star forming regions but for more than 1000 protostars [31]. Furthermore, multiple systems in the nearby regions, with separations as small as 3 au (at 230 pc), can also be detected if the circumstellar disk radii are ~1 au in radius [17, 18]. With such a facility and large samples, Science Ready Data Products will be essential.

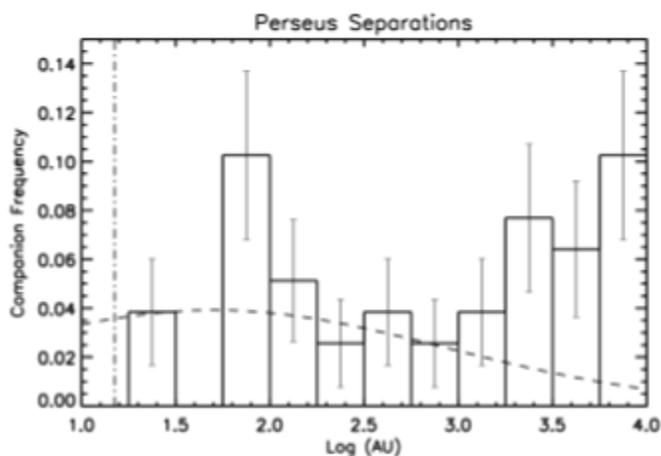

**Figure 2.** Histogram of protostar companion frequency versus separation in the Perseus cloud11]. The dashed curve is the same distribution but for the field solar-type stars [1]. The vertical dot-dashed line is the VLA resolution limit VLA (15 au),.

**Evolution of Protostellar Multiplicity**

Detecting large numbers of the earliest protostellar multiple star systems is just one piece required to understand multiple star formation. The evolution of multiplicity from the protostellar phase to pre-main sequence to field stars must also be understood. While we have emphasized the importance of a new mm/cm-wave facility, such a facility is not a panacea. As a young star system evolves from the protostellar phase, its surrounding envelope and dusty disk dissipates and eventually the individual circumstellar disks used to trace protostellar multiplicity will be too faint to detect at long wavelengths, even with a new mm/cm-wave facility.

In order to characterize multiplicity in protostellar systems evolving toward the pre-main sequence, ground-based near and mid-infrared facilities equipped with adaptive optics (AO) and space-based observatories (i.e., JWST) will be essential [19]. Young, forming star systems can often be obscured by high extinction at optical wavelengths, therefore to examine such systems, it is essential that AO systems on the current 8m class telescopes and future ELT-type facilities have near-infrared wavefront sensors to enable high resolution study of protostars. Near-infrared wavefront sensors are necessary, even when laser guide star AO is available, because there may not be an available optical star suitable for tip-tilt correction. Furthermore, an ELT-class telescope will be a crucial complement to a new mm/cm-wave facility, having nearly equivalent angular resolution, enabling multi-wavelength studies at the same resolution.

JWST in the near future will enable the detection and characterization of young multiple systems with separations as close as ~0.1" via medium resolution spectroscopy. This capability will enable the detection of photospheric features, accretion diagnostic spectral lines, and their relative flux densities. Photospheric features are needed to determine the spectral types of protostars in multiple systems, which are important to understand both the mass ratio of the companions, and accretion diagnostics will reveal how rapidly the protostars are gaining mass. The spectral types of the young stars can also be interpreted with respect to the kinematic mass measurements from the surrounding disk(s) to better understand protostellar evolution.

While new facilities with higher angular resolution are important, there is still a need for low to medium resolution spectroscopy in seeing-limited conditions. Throughput does not always increase with AO, and faint companions, with masses perhaps down to planetary masses, need to be characterized with broadband low resolution spectroscopy. Such capabilities are important for understanding the origin of wide, planetary-mass companions and brown dwarfs.

Finally, it is essential that efficient, high-resolution near-infrared spectroscopic instruments are available on the 8m-class and ELT-type facilities for long-term multi-object and multi-epoch campaigns. Protostars can be rapidly accreting material, such that the photospheric lines are heavily veiled, even in the near-infrared. Thus, high-resolution spectrometers are essential for measuring the spectral types of protostars [20]. Furthermore, high-dispersion near-infrared spectrographs can make it possible to search for spectroscopic binary protostars at younger evolutionary stages. This will enable proto-binaries to be discovered on scales smaller than a few au, allowing for a view into pathways of their formation and their environmental effects.

**Mass Ratios, Molecular Line Kinematics, and Magnetic Fields**

The masses of the primary stars (and secondaries when possible) also offer important clues to their origin and relation to the field star populations. For the more evolved protostars, as discussed previously, spectral types can be used to determine their masses. However, the masses of the youngest and deeply embedded protostars can only be determined from their protostellar disk kinematics (circumstellar or circumbinary). Facilities such as ALMA are now enabling these measurements with sensitive molecular line observations. However, more rarefied species such as $C^{17}O$ and $C^{18}O$ must be utilized to examine the protostellar disk kinematics through the surrounding envelope, increasing the amount of observing time required [21, 22]. Also, at short mm-wavelengths ($\geqslant$1.3 mm) the dust emission from protostellar disks is often optically thick [12], motivating observations at longer wavelengths (i.e., 3 mm) of both dust emission and molecular lines. However, the CO (1-0) isotopologue transitions at 3 mm are not as bright as the 1.3 mm (2-1) lines in protostellar disks that typically have higher temperatures. Thus a facility with greater sensitivity at 3 mm could enable superior measurements of protostellar mass because the dust emission will be less opaque than at the shorter wavelengths where ALMA is most efficient. See other WP by Tobin et al. for further discussion of protostellar mass measurements from molecular line kinematics. Often only a combined protostellar mass will be possible to measure from such kinematic observations, but the kinematic center, relative to the protostar positions, can then be used to estimate the companion mass ratio. For more evolved systems that can be resolved in the near-infrared, the spectral types of the two components can also be used to estimate the mass ratio, in addition to a mass ratio that may be derived from circumbinary disk kinematics.

The kinematic structure of multiple protostar systems from molecular lines can help reveal their formation mechanism. Systems can be examined for consistency with disk fragmentation (surrounding Keplerian disks), and wider systems with individual disks can have their relative velocities assessed (with respect to their protostar masses when available) to determine if the system is bound. Also, the relative disk orientations and outflow orientations can be examined to determine if ordered rotation likely formed the system or if turbulent fragmentation likely triggered the formation of the system [10]. Further development of polarimetric capabilities are important to enable the magnetic field structure and strength to be characterized in multiple systems. This will enable the role of magnetic fields in the formation and evolution of binary systems to be fully assessed and whether they may affect the companion separations [30].

**Multiplicity in Context**

If more distant young star systems are going to be examined by a new mm/cm-wave facility and upcoming ELT-type facilities, it is essential to characterize the protostellar luminosities and evolutionary states. *Spitzer* and *Herschel* were essential for this in the nearby star-forming regions, but source confusion will be a problem for more distant star forming regions. JWST will enable the best angular resolution thus far in the mid-infrared. This will reduce confusion in distant star-forming regions, and total luminosities can be estimated from the JWST spectral range [23], but JWST has slow mapping speed. Accurately characterizing the SEDs of distant protostars requires mid-to-far-infrared facilities operating between ~3.6 - 250 µm with a larger apertures than *Spitzer*/*Herschel* and better survey speed than JWST. A key strength of the previous VLA protostellar multiplicity surveys has been the availability of well-characterized samples of protostars, and complementary facilities will be essential to understand the context of multiplicity.

**Theoretical Advances and Model Development**

From a theoretical perspective, the last decade has brought great strides in modelling protostellar multiplicity via the investigation of different physical processes, including treatment of stellar feedback [24,25,26]. However, the origin of multiplity and its dependence on both the local and global star-forming environment remain debated. Hydrodynamic efforts suffer from the same limitations as observations: poor resolution on ~< au scales, small number statistics ($N^* < 100$), and incomplete exploration of high-mass and extreme star-forming environments. Meanwhile, uncertainty persists in the dynamical state and conditions of molecular clouds, which influence the characteristics of multiplicity on smaller scales [6, 25]. Close comparison between theoretical models and observations, "synthetic observations," are required to close the loop and are now being carried out with increasing regularity [16, 27, 24, 28].

During the next decade, progress is required on two fronts. First, multi-physics treatments must include the adoption of non-ideal magnetic effects and astrochemistry, which are required to accurately model protostellar disk characteristics, microphysics and stability. Realistic abundances and gas temperatures are also prerequisites for producing accurate synthetic observations and thus interpreting observations and discriminating between models. Second, the continued expansion of high-performance computing facilities is essential to enable calculations with the necessary resolution, statistics and physical conditions. Fostering close, interdisciplinary collaboration with computer science will pave the way to developing more efficient and innovative algorithms. Increasing access to facilities, as well as publicly disseminating codes, will ensure participation by a wider segment of the community.

**Summary**

Studies of protostellar multiplicity have revolutionized our knowledge of multiple star formation in the past decade through surveys at infrared and mm/cm wavelengths. However, the limits of the current instruments are being reached and in order to fully understand the origin of multiple stars, a new mm/cm-wave facility is required. Such a facility will be made significantly more powerful with complementary advances in ground/space-based near to far-infrared imaging and near to mid-infrared spectroscopy. Such facilities will open up new, exciting regions of parameter space in protostellar multiplicity studies. Companions with ≳5× smaller separations will be able to be examined and larger samples of low-mass protostars within more distant regions (e.g., Cygnus-X) can be examined. Theoretical advances in tandem with observational progress are essential for a complete understanding of multiplicity at scales and evolutionary times that will continue to remain inaccessible to observations.